\documentclass[twocolumn,showpacs,nofootinbib]{revtex4-1}

\usepackage{amsmath}
\usepackage{graphicx}
\usepackage{dcolumn}
\usepackage{bm}
\def\to{\rightarrow}

 \def\CQG{{\it Class. Quantum
		Gravity} }

 \def\PL{{\it Phys. Lett.} } \def\PR{{\it
		Phys. Rev.} } \def\PRL{{\it Phys. Rev. Lett.} } \def\PRTS{{\it
		Physics Reports} }

  \def\ga{\gamma}

  \def\mn{{\mu\nu}}

 \def\frac#1#2{{\textstyle{{#1}\over
			{#2}}}} 
\def\lsim{\mathrel{\rlap{\lower4pt\hbox{\hskip1pt$\sim$}}
		\raise1pt\hbox{$<$}}}
\def\gsim{\mathrel{\rlap{\lower4pt\hbox{\hskip1pt$\sim$}}
		\raise1pt\hbox{$>$}}} \def\sqr#1#2{{\vcenter{\vbox{\hrule
				height.#2pt \hbox{\vrule width.#2pt height#1pt \kern#1pt \vrule
					width.#2pt} \hrule height.#2pt}}}}
\def\square{\mathchoice\sqr66\sqr66\sqr{2.1}3\sqr{1.5}3}
\def\beq{\begin{equation}} \def\eeq{\end{equation}}
\def\beqa{\begin{eqnarray}} \def\eeqa{\end{eqnarray}}

\usepackage{hyperref}

\begin{document}

\preprint{APS/123-QED}

\title{Inflation with a massive vector field nonminimally coupled to gravity}

\author{O. Bertolami}
\email[]{orfeu.bertolami@fc.up.pt}
\author{V. Bessa}
\email[]{up201000387@fc.up.pt}
\author{J. P\'aramos}
\email[]{jorge.paramos@fc.up.pt}
\affiliation{Departamento de F\'\i sica e Astronomia and Centro de F\'isica do Porto, \\Faculdade de Ci\^encias da Universidade do Porto,\\Rua do Campo Alegre 687, 4169-007 Porto, Portugal}

\date{\today}

\begin{abstract}
We study the possibility that inflation is driven by a massive vector field with $SO(3)$ global symmetry nonminimally coupled to gravity. From an $E^3$-invariant Robertson-Walker metric we propose an \textit{Ansatz} for the vector field, allowing us to study the evolution of the system. We study the behaviour of the equations of motion using the methods of the theory of dynamical systems and find exponential inflationary regimes. 

\end{abstract}


\maketitle

\section{\label{sec:level1}Introduction}

Inflation is a key ingredient for solving the initial conditions problems of the Cosmological Standard Model \cite{Starobinsky,Guth,Albrecht,Linde} (see {\it e.g.} Ref. \cite{Olive} for a review). Inflation is most often driven by a scalar field  with a suitable potential; however, it would be particularly interesting that if inflation could also be set up with vector fields. The issue has been considered long ago in Refs. \cite{Ford,artigo}, resorting to a quadratic potential $ V(A^\mu) \sim A_\mu A^\mu $, and more recently in Ref. \cite{Mota} (see also Refs. \cite{inflation1,inflation2,inflation3,inflation4}). It was shown in Ref. \cite{artigo} that a massive vector field cannot, on its own, drive inflation.

The purpose of this paper is to investigate the possibility of an inflationary regime using a massive vector field, if the latter is nonminimally coupled to gravity. We shall consider the proposal put forward in Ref. \cite{artigo}, by further including a nonminimal coupling between the vector field and both the Ricci scalar and Ricci tensor. Indeed, from a phenomenological point of view, models containing only a coupling to the Ricci scalar can be viewed as incomplete or unnatural, as there is no {\it a priori} reason for assuming that a coupling of the vector field to the Ricci tensor is negligible when compared with the latter.

The first coupling has been examined as a mechanism to drive inflation in Refs. \cite{inflation2,inflation4}, albeit with an {\it Ansatz} for the behaviour of the vector field quite different from the one assumed in this study; it has also been approached in the context of primordial magnetic field generation \cite{Mota2} (see also Ref. \cite{Turner}). The second type of coupling has been considered in models of gravity with spontaneous breaking of Lorentz symmetry \cite{bumblebee1,bumblebee2,bumblebee3}.

The inclusion of only a coupling $\sim A_\mu A^\mu R$ can be interpreted as a varying mass term for the vector field, which breaks its $U(1)$ symmetry and induces the appearance of a longitudinal mode absent in massless vector fields (such as the photon).

As Ref. \cite{ghost1} has shown, albeit with rather different assumptions for the vector field (namely the lack of $SO(3)$ invariance leading to the {\it Ansatz} described in the next section), computing the perturbative spectrum of the model reveals that such a mode is a ghost: expanding it around a Friedmann-Robertson-Walker (FRW) shows that the corresponding perturbation is endowed with a kinetic term of the wrong sign. The addition of a ``hard'' mass $m^2$ cannot alleviate this problem, as the effective mass will nevertheless shift sign after inflation ends \cite{ghost1,ghost2}.

Naturally, the equations for cosmological perturbations for the model here considered are much more convoluted, and approaching them is beyond the purported scope of this work: however, it is possible that the additional dynamics could avoid the appearance of ghost modes described above. In particular, it is worth pointing out that a coupling $\sim R_\mn A^\mu A^\nu$ can be interpreted as an additional contribution to a mass term during De Sitter inflation --- since, for a constant expansion rate, $R_\mn \sim g_\mn R$ and both couplings are similar, so that their effect is dynamically equivalent (as shall be shown later); however, this is no longer the case for a realistic inflationary scenario, where the expansion of the Universe is not perfectly exponential.

This work is organised as follows: in Section 2 we present our model and the corresponding field equations; the ensuing cosmological dynamics is discussed in Section 3, focusing on the possibility of exponential inflation and some specific regimes allowing its occurrence. Section 4 presents the dynamical system derived from the cosmological equations, derives its critical points and the corresponding values for the expansion rate. Finally, in Section 5 we present our conclusions.

\section{\label{sec:levelII}The Model}

We  consider the action for an $SO(3)$-invariant gauge group with a massive vector field nonminimally coupled to the curvature:

\begin{equation}\label{action}
S = \int d^4x \sqrt{-g} \left({R \over 2k^2}+\mathcal{L}\right)~~,
\end{equation}

\noindent with

\begin{eqnarray}\label{Lagrangedensity}\mathcal{L} &=& {1 \over 8e^2}Tr[F_{\mu\nu}F^{\mu \nu}]+{1\over 2}m^2Tr[A_\mu A^\mu]+ \nonumber \\
&& {1 \over 3} \alpha RA_\mu A^\mu + \beta R_{\mu \nu}A^\mu A^\nu ~~,\end{eqnarray}

\noindent where $k^2=8\pi G$, $e$ is the gauge coupling, $ \alpha $ and $ \beta $  are the strengths of the nonminimal couplings between the gauge field and the Ricci scalar and Ricci Tensor, respectively \cite{artigo}. The gauge field strength is given by $F_{\mu\nu}=\partial_\mu A_\nu - \partial_\nu A_\mu + [A_\mu , A_\nu]$.

\subsection{\label{Sec:levelA} Field Equations}
The variation of the action with respect to the metric yields:

\begin{eqnarray}\label{Eeq}
&&{1 \over 2k}G_{\mu \nu} = -{1 \over k^2}R_{\mu \nu} + g_{\mu \nu}\mathcal{L}- m^2 A_\mu A_\nu  -{1 \over 2e^2}Tr[F_{\mu \nu}F^{\mu \nu}] \nonumber \\ && -{2\over 3}\alpha\big[R_{\mu \nu}(A^\rho_\rho)+RA_\mu A^\nu -\nabla_\mu \nabla_\nu (A_\rho A^\rho) + g_{\mu \nu}(A_\rho A^\rho)\big] \nonumber \\ \nonumber
&& +  \beta\big[2\nabla_{\beta (\mu}A_{\nu)}A^\beta -g_{\mu \nu}(\nabla_\alpha \nabla_\beta A^\alpha A^\beta) \\ &&-\square (A_\mu A_\nu)-4A^\alpha R_{\alpha(\mu}A_{\nu)}\big]~~,
\end{eqnarray}

\noindent where $G_{\mu\nu}$ is the Einstein tensor.

Variation with respect to the gauge field yields the vector field equations of motion:

\begin{equation}\label{fieldeq}
{1\over 8e^2} \nabla_\mu (\nabla_\nu A^\nu) + \left( {1\over 2}m^2 + {1 \over 3}\alpha R\right) A_\mu + \beta R_{\mu \nu}A^\nu = 0 ~~.
\end{equation}

\section{\label{sec:levelIII} Cosmological Dynamics}

In this work we use the $SO(3)$-invariant {\it Ansatz} discussed in Ref. \cite{artigo}, which is briefly reviewed here. The geometry associated with the flat FRW universe has the form $M^4=R^4=R \times E^3/SO(3)$, where $E^3$ represents a six-dimensional Euclidean group of spacial hypersurfaces. Compatibility with the FRW geometry requires that the vector field is an $SO(3)$-invariant multiplet $A_\mu^a$, $a=1,...,N$, where $a$ is an internal index.

The Robertson-Walker metric has the form:

\begin{equation}\label{metric}
ds^2=-N(t)^2(dx^0)^2+a(t)^2\sum\limits_{i=1}^{3}(dx^i)^2~~,
\end{equation}

\noindent where $N(t)$ is the lapse function and $a(t)$ is the scale factor. If the lapse function is set to identity, $N(t) = 1$, the (diagonal) Ricci tensor and ensuing Ricci scalar read

\begin{eqnarray}
&& R_{tt}= -3 {\ddot{a}\over a} ~~, \nonumber \\
&& R_{ii}=  2 (\dot{a})^2 + a \ddot{a} ~~~~\to~~~~ R= 6\left[ \left({\dot{a}\over a}\right)^2 + {\ddot{a}\over a}\right]~~.
\end{eqnarray}

Imposing spatial homogeneity and isotropy, it is found that the vector field must have the following form \cite{artigo}:

\begin{equation}\label{ansatz}
A_0 = 0~~~~,~~~~A_i= A^a_i L_a=\chi_0(t) \delta^a_i L_a~~,
\end{equation}

\noindent where $\chi_0(t)$ is an arbitrary function of the time and $L_a$ are the generators of the internal $SO(3)$ group.

The cosmological field equations follow from the substitution of Eqs. (\ref{metric})-(\ref{ansatz}) into Eqs. (\ref{Eeq}) and (\ref{fieldeq}): these equations can be more promptly found from the effective action, since all constraints are respected; the former is obtained by replacing the {\it Ansatze} Eqs. (\ref{metric}) and (\ref{ansatz}) into action Eqs. (\ref{action})-(\ref{Lagrangedensity}) and discarding the infinite volume of the spatial hypersurface, yielding:

\begin{eqnarray}\label{action2}
S_{\rm eff} = 3\int dt &\Bigg[&-{a \dot{a}^2 \over k^2N} + {a\over 4Ne^2}\left({\dot{\chi}_0^2\over 2}-{N^2\over a^2}V(\chi_0)\right) + \nonumber \\ 
&& \left({1\over 4}Nm^2 + \gamma{\dot{a}^2\over Na}\right)\chi_0^2\Bigg] ~~,
\end{eqnarray}

\noindent where we define the quartic potential $V (\chi_0) = \chi_0^4/8$ and the composite coupling $\gamma \equiv \alpha - \beta$, showing that the contribution of the two couplings between the vector field and the curvature have similar dynamical impact; interestingly, the case when $\beta =\alpha$ yields a vanishing effect from to the aforementioned couplings. In the remainder of this study, we thus consider that $\gamma \neq 0$.

Varying the above with respect to $a(t)$, $N(t)$ and $ \chi_0(t)$ and setting the gauge $N(t)=1$, we get the Friedmann and Raychaudhuri equations, together with the equation of motion for the vector field:

\begin{eqnarray} \label{friedmann}
&& 4(a^2-k^2\gamma\chi_0^2)H^2 = {k^2\over e^2}\left({\dot{\chi}_0^2\over 2}+{V \over a^2}\right)+ k^2m^2\chi_0^2~~,\\ \label{raychaudhuri}
&& (a^2-k^2\gamma\chi_0^2)(\dot{H}+H^2) = \nonumber \\ && -H^2 a^2+  k^2\left( 2\gamma\dot{\chi}_0\chi_0H+{m^2\chi_0^2\over 4}\right)~~, \\ \label{motion} &&\ddot{\chi}_0 + H\dot{\chi}_0 = -{\chi_0^3\over 2a^2}+ 8e^2H^2\gamma\chi_0- 2e^2m^2\chi_0~~,
\end{eqnarray}

\noindent where $H=\dot{a}(t)/a(t)$ is the expansion rate.

\subsection{\label{sec:levelV}De Sitter Phase}

Before a more encompassing study of the dynamical system resulting from the above Eqs. (\ref{friedmann})-(\ref{motion}), we may first ascertain wether a solution with an exponential scale factor solution is admissible:

\begin{equation}
a(t)\sim e^{H_0t}~~,
\end{equation}

\noindent where $H_0$ is a constant and $t $ the cosmic time. Since $H(t) = H_0$, setting $\dot{H}=0$ in Eqs. (\ref{friedmann})-(\ref{motion}), together with the {\it Ansatz} $\chi_0(t) = A a(t)$, yields

\begin{eqnarray}
4(1-k^2\gamma A^2)H_0^2 &=& k^2A^2\left[ { H_0^2\over 2}+{A^2\over 8} + m^2 \right] ~~, \nonumber \\
4(2 - 3 k^2\gamma A^2)H_0^2 &=& k^2 A^2 m^2 ~~,\nonumber \\ 4 ( 1 - 4 \gamma ) H_0^2 &=& -A^2 - 4 m^2~~,
\end{eqnarray}

\noindent where we have fixed $e=1$, for brevity. These equations have as solutions

\begin{eqnarray}\label{deSitter}
H_{0\pm}^2 & = & {2 +(1+8\gamma) (mk)^2  \over 24 k^2 \gamma (4\gamma - 1 )} \times \nonumber \\ && \left( 1 \pm \sqrt{1 + {48(4\gamma-1)\ga (mk)^4 \over [2 + (1+8\gamma) (mk)^2]^2 }} \right) ~~,
\end{eqnarray}

\noindent and

\begin{equation}A^2= 4 \left[ ( 4 \gamma - 1 ) H_0^2 - m^2 \right] = { 8 \over k^2 \left[12 \gamma + \left({m\over H_0}\right)^2 \right] } ~~.
\end{equation}

For $\chi_0(t)$ to be real we must have 

\begin{equation} \label{conditions} ( 4 \gamma - 1 ) H_0^2 > m^2 ~~~~,~~~~12\gamma H_0^2 + m^2 > 0~~. \end{equation}

\noindent These conditions, together with the requirement of a real expansion rate $H_0^2 > 0$, imply that only the positive branch $H_{0+}$ should be considered, and that the coupling strength must obey the restriction $\gamma > 1/4$.

\subsubsection{\label{sec:levelC} Massless case}

In the massless case, $m=0$, the above Eqs. turn to:

\begin{equation}
H_0^2 = {1 \over 6 k^2 \gamma (4\gamma - 1 )} ~~~~,~~~~\chi_0(t) = \sqrt{ 2 \over 3 \gamma  }{a(t)\over k} ~~.
\end{equation}

\subsubsection{\label{sec:levelD} Strong coupling limit}

In the strong coupling limit $ (mk)^2\gamma \gg 1$, we can perform a first order expansion of the Eqs. (\ref{friedmann})-(\ref{motion}), obtaining:

\begin{equation}
H_0 = \sqrt{1 \pm 2 \over 12 \gamma }m = \begin{cases}  &{m \over 2 \sqrt{\gamma} }~~,~~ \gamma >0 \\ &{m \over 2\sqrt{-3 \gamma} }~~,~~ \gamma <0 \end{cases} ~~.\end{equation}

However, while the positive coupling case $\gamma >0$ yields a real valued vector field, with

\begin{equation} \chi_0(t) = {1 \over \sqrt{2\gamma}}{a(t) \over k }~~,\end{equation}

\noindent the converse case, $\gamma <0$, yields an imaginary function

\begin{equation} \chi_0(t) = {4 i\over \sqrt{3}} m a(t)~~, \end{equation}

\noindent indicating that a strong coupling is only viable if the coupling to the Ricci scalar is stronger than to the Ricci tensor, $\alpha > \beta$.

\subsubsection{\label{sec:levelE} Weak coupling limit}

In the weak coupling limit $(mk)^2\gamma \ll 1$ and $m\neq 0$, we obtain to first order of Eqs. (\ref{friedmann})-(\ref{motion}) only one real solution,

\begin{equation}
H_0 = {1 \over 2k} \sqrt{2+(mk)^2\over -3\gamma}~~~~,~~~~\gamma <0~~,
\end{equation}

\noindent which requires that $\gamma <0$. However, this also leads to 

\begin{equation} \chi_0(t) = \sqrt{2+(mk)^2 \over 3\gamma} {a(t) \over k}~~,
\end{equation}

\noindent which is thus imaginary --- as expected, since it breaks the previously obtained condition $\gamma > 1/4$. As such, we conclude that no weak coupling regime is possible with a massive vector field.

Before proceeding, we may also check if a power-law behaviour for the scale factor and vector field is viable: by setting $a(t) \sim t^p $ and $\chi_0 \sim t^n$, we obtain $H(t) \sim t^{-1}$, and from Eqs. (\ref{friedmann})-(\ref{motion}), the relationships:

\begin{eqnarray}
0&=& At^{2p-2}+Bt^{2n-2} + Ct^{4n-2p} + Dt^{2n}~~, \\ \nonumber 
0&=& Et^{2p-2}+Ft^{2n-2} + Gt^{2n} ~~, \\ \nonumber 0&=& Ht^{n-2} + It^{3n-2p} + Jt^n~~,
\end{eqnarray}

\noindent where the capital letters denote non-vanishing constants. Thus, it is clear that one cannot find a simple monomial solution for the dynamical system ensued by Eqs. (\ref{friedmann})-(\ref{motion}), further motivating a rigorous study of its critical points.

\section{\label{sec:levelVIII} Dynamical System}

To find general inflationary solutions, we must solve the dynamical system associated with Eqs. (\ref{friedmann})-(\ref{motion}) and physically interpret the ensuing critical points. To do so, the following dimensionless variables are introduced,

\begin{eqnarray}\label{def}
x = {k\chi_0(t)\over a(t)\sqrt{1-w^2}} ~~~~&,&~~~~
y = {k^2\dot{\chi}_0(t) \over 2 \sqrt{2(1-w^2)}} ~~,\\ \nonumber
z = kH ~~~~&,&~~~~ \tau = {t \over k}~~,
\end{eqnarray}

\noindent where the auxiliary function 

\begin{equation}
w=\sqrt{\gamma} {k\chi_0(t) \over a(t)} = x \sqrt{\gamma \over  1+ \gamma x^2}~~,
\end{equation}

\noindent is defined for convenience.

The Friedmann Eq. (\ref{friedmann}) yields an algebraic constraint:

\begin{equation}\label{constraint}
z^2=y^2+{1\over 32}{x^4 \over 1+\gamma x^2}+{1\over 4}\mu^2x^2~~,
\end{equation}

\noindent where $\mu = mk$ is the reduced mass, and only two degrees of freedom remain. We can now calculate the derivative of the variables $(x,y)$ with respect to the dimensionless time $\tau$, obtaining

\begin{eqnarray}
\nonumber x_\tau \equiv {d x \over d\tau} &=&  (1+\gamma x^2) \left[ y-x \sqrt{y^2+{1-w^2\over 32}x^4+{1\over 4}\mu^2x^2} \right] ~~,\\ \nonumber y_\tau \equiv {d y \over d\tau} &=& -{1-w^2\over 4\sqrt{2}}x^3-  {\mu^2\over \sqrt{2}}x \\ \nonumber && +{4\gamma\over \sqrt{2}}\left(2y^2+{1-w^2\over 32}x^4+{1\over 4}\mu^2x^2\right)x-\nonumber \\ \label{dynsys}
&& (\gamma x^2+2)y\sqrt{y^2+{1-w^2\over 32}x^4+{1\over 4}\mu^2x^2} ~~.
\end{eqnarray}

\subsection{Finite critical points}

Considering the dynamical system Eq. (\ref{dynsys}), we first analyse the  the origin $F(0,0)$, a trivial critical point. The eigenvalues of the Jacobian matrix derived from Eq. (\ref{dynsys}) are $\lambda_\pm = \pm 2i\sqrt{2}\mu$, indicating that this is a stable critical point; this is a natural result, as the vector field vanishes and as such the nonminimal couplings have no impact, collapsing to the case studied in Ref. \cite{artigo}.

Asides from the above, eight non-trivial critical points arise, as shown in Table 1. For convenience, we define

\begin{eqnarray} X_\pm &=& { 2 + \mu ^2 \pm \sqrt{(1-16 \gamma )^2 \mu ^4+4 (8 \gamma +1) \mu ^2+4} \over 2 \gamma \left[ 1 + (8 \gamma -1) \mu ^2 \right]} ~~,  \nonumber \\ Y_\pm &=& {1 \over 12\gamma(4\gamma-1)} \left[ \mu^2 + { 2 + (1+4\gamma)\mu ^2 \over 8 } X_\pm \right] ~~.\end{eqnarray}

In what follows it is relevant to realise that the dynamical system is invariant under reflections $(x,y)\to (-x,-y)$, so that it suffices to analyse the first four critical points (first column).

Notice that the value for the expansion rate, read from the algebraic constraint Eq. (\ref{constraint}), naturally coincides with those discussed in the previous section, as can be seen from Eq. (\ref{deSitter}). This also indicates that the critical points $(C,D,G,H)$ are unphysical, as they lead to an imaginary value $H_-$ for the expansion rate.
\begin{table}[!h]
\label{fixed1}
\centering
\begin{tabular}{ccc|ccc}
Point & $(x,y)$ & $H $ & Point & $(x,y,z)$ & $H$ \\ \hline
A & $(\sqrt{X_+},\sqrt{Y_+})$ & $H_+$ & E & $(-\sqrt{X_+},-\sqrt{Y_+})$ & $H_+$ \\ \hline
B & $(\sqrt{X_+},-\sqrt{Y_+})$ & $H_+$ & F & $(-\sqrt{X_+},\sqrt{Y_+})$ & $H_+$ \\ \hline
C & $(\sqrt{X_-},\sqrt{Y_-})$ & $H_-$ & G & $(-\sqrt{X_-},-\sqrt{Y_-})$ & $H_-$ \\ \hline
D & $(\sqrt{X_-},-\sqrt{Y_+})$ & $H_-$ & H & $(-\sqrt{X_-},\sqrt{Y_-})$ & $H_-$

\end{tabular}
\caption{Non-trivial, finite critical points.}
\end{table}

Given the convoluted expressions for the eigenvalues of the Jacobian matrix evaluated at the critical points $(A,B,C,D)$, we follow a numerical procedure to show that these are saddle points.

For this, we assign a range of values for the coupling $\gamma$ and the reduced mass $\mu$ and numerically compute the value of the real part of the two eigenvalues  $\lambda_1$ and $\lambda_2$ of the Jacobian matrix. We find that the Jacobian evaluated at the critical points $(A,B)$ has the same two eigenvalues $\lambda_1$ and $\lambda_2$, with real parts that are almost symmetric; the same behaviour occurs for the pair $(C,D)$. This is graphically shown in Figs. 1 and 2, where it is clear that these critical points fall neatly in the line $Re(\lambda_1) = -Re(\lambda_2)$.

\begin{figure}
	\label{real1}
	\centering
	\includegraphics[height=8cm]{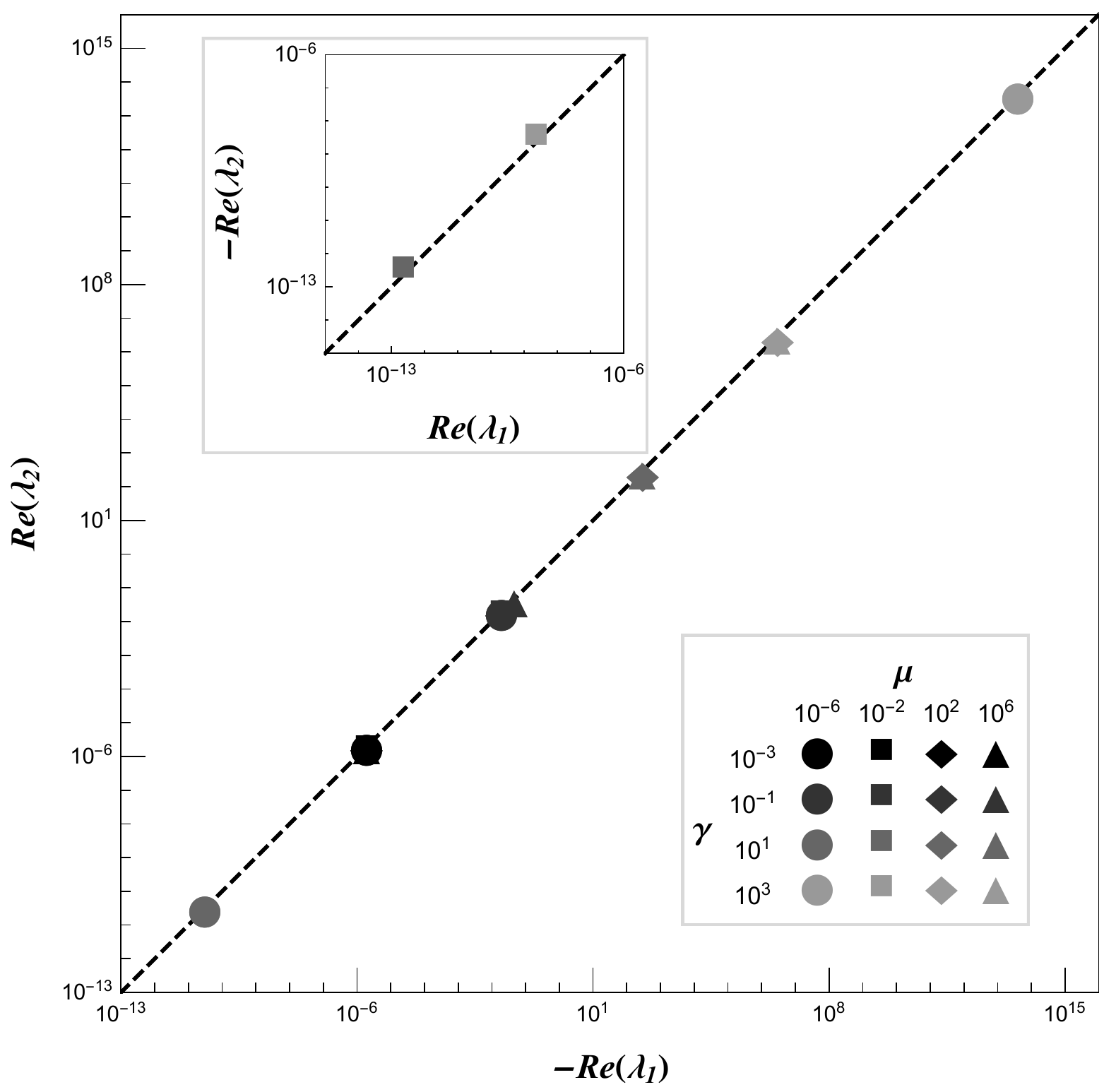}
	\caption{Real part of the eigenvalues of critical points $(A,B)$.}
\end{figure}

\begin{figure}
	\includegraphics[height=8cm]{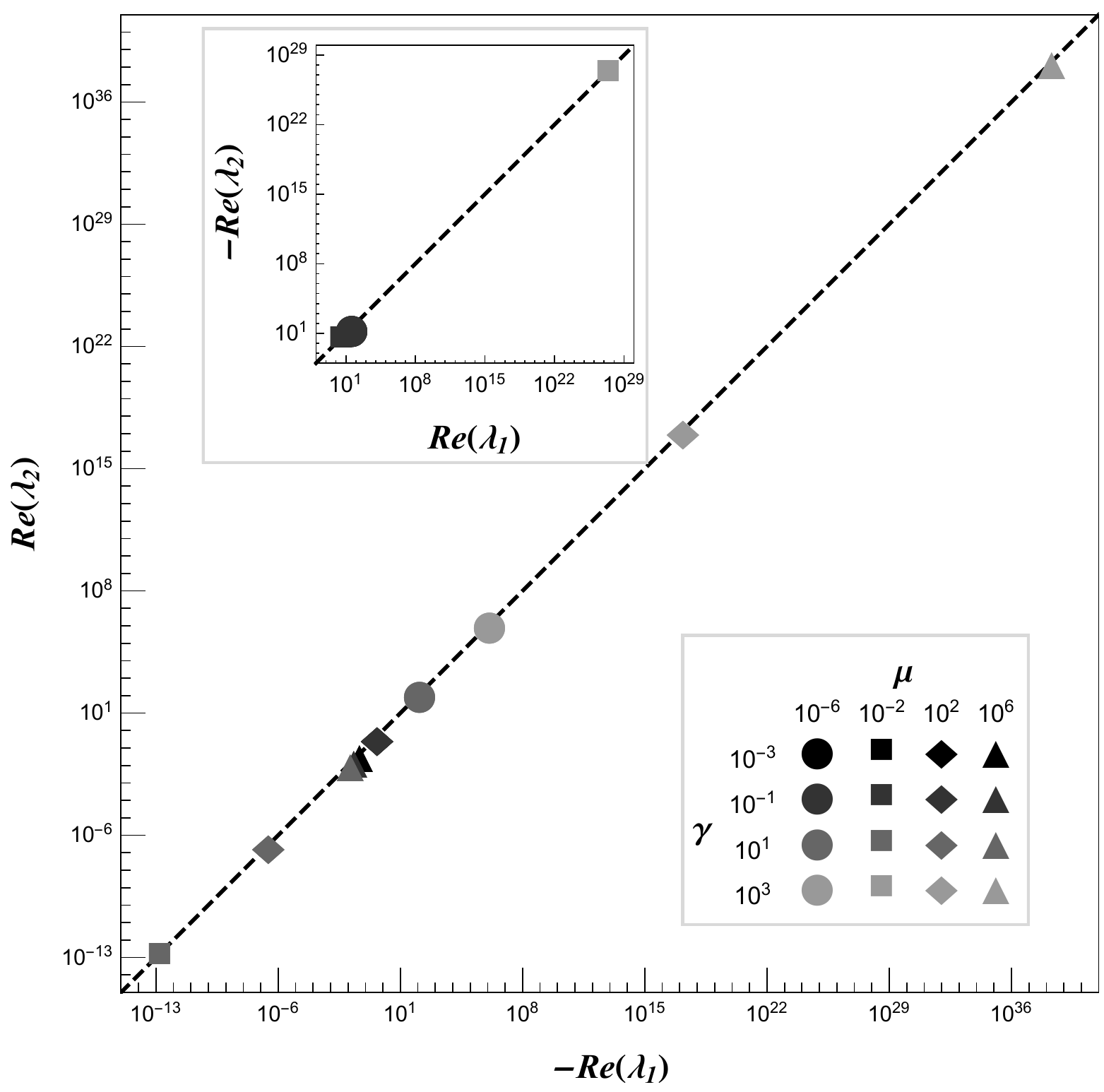}
	\caption{Real part of the eigenvalues of critical points $(C,D)$.}
\end{figure}

We notice that it does not suffice to require that $(x,y)$ are real-valued for the critical points $(A,B)$ to be physically meaningful: indeed, one must also consider the definition of the dimensionless variables Eq. (\ref{def}) to ensure that the physical vector field and its time derivative are well defined. This requires that $1-w^2 >0 \to 1+ \gamma x^2 >0$, which together with the requirement that $(x,y)$ are real translates into condition $ \gamma >1/4$ --- precisely the constraint obtained in the previous section.

\subsection{\label{sec:levelB}Critical points at infinity}

We now analyse putative critical points found at infinity, by resorting to a boundary at infinity, $x^2+y^2 = \infty$, which is then compactified to a circle of unit radius. In order to do so, we introduce a new radial coordinate and time variable, together with the usual definition of polar angle, through:

\begin{equation}
	x = {\rho \cos \theta\over 1-\rho}~~,~~y = {\rho\sin \theta\over 1-\rho}~~,~~{d\zeta \over d\tau} = {1 \over (1-\rho)^2}, \end{equation}

\noindent where $0\leq \rho \leq 1$ .

The dynamical system Eq. (\ref{dynsys}) may be rewritten as

\begin{widetext}
\begin{eqnarray}\label{dynsys2}
\rho_\zeta \equiv {d \rho \over d\zeta} = \Pi(\rho,\theta) &=& {1\over 2}\big[-3 +6\rho - (3+\gamma)\rho^2 + [1-2\rho+(1-\gamma)\rho^2] \cos 2\theta \big] \rho^2 f(\rho,\theta) + {\sqrt{2} \over 16} { \rho g(\rho,\theta) \sin \theta \cos \theta \over (1-\rho)^2 + \gamma \rho^2 \cos^2 \theta } ~~, \nonumber \\ \theta_\zeta \equiv {d \theta \over d\zeta} = \Psi(\rho,\theta) &=& - \rho f(\rho,\theta) \sin \theta \cos \theta + {\sqrt{2} \over 16} {h(\rho,\theta) \over (1-\rho)^2 + \gamma \rho^2 \cos^2 \theta } ~~,
\end{eqnarray} 
\noindent with
\begin{eqnarray}
f^2(\rho,\theta) &=& (1-\rho) \left( {\rho ^2 \cos ^4\theta \over 32 \left[(1-\rho)^2 + \gamma  \rho ^2 \cos ^2\theta \right]}+ {1\over 4} \mu ^2 \cos ^2\theta + \sin ^2\theta \right) ~~, \\ \nonumber 
g(\rho,\theta)&=& 8 (\mu^2 - 4 )  ( 1 -5 \rho)  + \big( 16 [5 \mu^2 -4\gamma -10] + 2\cos^2\theta \big)\rho^2 + 2\big ( 160 + 96\gamma -40 \mu^2 - 3 \cos^2 \theta \big) \rho^3 + \\ \nonumber && \big( 8 [5 \mu^2 - 24 \gamma - 20 ] + 2[3 -32 \gamma^2] \cos^2 \theta + [8\gamma( 4-\mu^2) - 1 ]\cos^4\theta \big) \rho^4 + \\ \nonumber && \big( 8 [4+8\gamma - \mu^2] - 2 [1 - 32 \gamma^2 ] \cos^2 \theta - [ 8\gamma ( 4- \mu^2 ) - 1 ] \cos^4 \theta  \big) \rho^5~~, \\ \nonumber h(\rho,\theta) &=& 8 ( 1 - \rho)^4 [ (4-\mu^2) \cos ^2\theta -4] - 2 \rho^2 [1- 2 \rho + (1-16\gamma^2)\rho^2 ] \cos^4 \theta+ \gamma [1 + 8\gamma (\mu^2 - 4) ] \rho^4 \cos^6 \theta~~.
\end{eqnarray}
\end{widetext}

The critical points at an infinitely distant boundary are obtained by finding the solution of the equations $\Pi(1,\theta)=0$ and $\Psi(1,\theta)=0$. Since Eq. (\ref{dynsys2}) is invariant under transformations $\theta \to \theta+\pi$, it suffices to consider just the critical points lying on the region $[0,\pi]$. As $f(1,\theta) = g(1,\theta)=0$ and $h(1,\theta) $, the critical points are given by

\begin{equation} \big[32 \gamma + (1+8\gamma [\mu^2 - 4]) \cos^2 \theta \big] \cos^2 \theta =0 ~~, \end{equation}

\noindent with solutions labelled as $N\left(1,{\pi \over 2}\right)$ and

$$S_\pm\left(1, \arccos\left(\pm \sqrt{1 \over 1-{1\over 32\gamma}-{1 \over 4} \mu^2 }\right) \right)~~.$$

The argument of the critical points $S_\pm$ is real under the condition

\begin{eqnarray} \label{constraintSpm1}
&& \left[ \mu <2 \wedge \left( \gamma <0 \vee \gamma > {1 \over 8(4-\mu^2)}\right)\right] \vee \nonumber \\
&& \left[ \mu > 2 \wedge -{1 \over 8(\mu^2 - 4)} < \gamma <0 \right]~~, \end{eqnarray}

\noindent where the symbols $\wedge$ (and) and $\vee$ (or) have been used.

The linearisation of the system Eq. (\ref{dynsys2}) allows for the derivation of the eigenvalues of the Jacobian around the critical points $S_\pm$ (which are degenerate) and $N$; in the latter case, this further requires a change in the time variable $\zeta \rightarrow \hat{\zeta}$, such that $d\hat{\zeta}/d\zeta=\rho-1$. The ensuing results are presented in Table 2.

\begin{table}[!h]
	\label{fixed2}
	\centering
	\begin{tabular}{cc}
		Point & Eigenvalues \\ \hline
		$ S_+$ & ${8 \gamma  \sqrt{-\gamma  \left(8 \gamma  \mu ^2+1\right)}\over 1 + 8 \gamma  \left(\mu ^2-4\right)}$ \\ $ S_-$ & $-{8 \gamma  \sqrt{-\gamma  \left(8 \gamma  \mu ^2+1\right)}\over 1 + 8 \gamma  \left(\mu ^2-4\right)}$ \\ $N$ & ${3 \pm \sqrt{1-64\gamma}\over 2}$                                                        
	\end{tabular}
	\caption{Eigenvalues of the critical points at infinity $S_\pm$ and $N$ for the dynamical system Eq. (\ref{dynsys2}). }
\end{table}

In order to extract the expansion rate, we again resort to the definition Eq. (\ref{def}) and the algebraic constraint Eq. (\ref{constraint}) which, in the compactified polar coordinates, reads

\begin{eqnarray} && \label{zsq} (kH)^2 = z^2(\rho,\theta) = {\rho ^2 \over 32 (1-\rho )^2} \times \\ \nonumber && \left({\rho ^2 \cos ^4\theta\over \gamma \rho^2 \cos ^2\theta+(1-\rho )^2}+ 8 \mu ^2 \cos ^2\theta+32 \sin ^2\theta \right) ~~.\end{eqnarray}

\subsubsection{Critical Point $N$}

Inspection of Table 2 shows that the critical point $N(1,\pi/2)$ is

\begin{itemize}
	\item a saddle point, if $\gamma \leq -1/8$;
	\item unstable, if $-1/8<\gamma <1/64$;
	\item a focus, if $\gamma > 64$.
\end{itemize} 

Replacing $\theta=\pi/2$ in Eq. (\ref{zsq}), we see that $z \sim 1/(1-\rho) \to \infty$ for all values of the coupling $\gamma$ and reduced mass $\mu$, thus yielding the possibility of a Big Rip scenario (if $N$ is a focus), {\it i.e.} the Universe evolves towards an infinite expansion rate.

\subsubsection{Critical Points $S_\pm$}

By inspecting Table 2, we may ascertain the behaviour of the critical points $S_\pm$: we impose condition Eq. (\ref{constraintSpm1}) for real critical points and vary the coupling $\gamma$ and reduced mass $\mu$ to determine the behaviour of the corresponding degenerate eigenvalues, as depicted in Fig 3. We find that the latter are never positive, yielding

\begin{itemize}
	\item $\gamma < 0 $: $-{1\over 8\mu^2} < \gamma <0 $ 
	\item $Re (\gamma ) = 0$: $\begin{cases} \gamma < -{1\over 8\mu^2} \vee \gamma > {1 \over 8(4-\mu^2)}~~,&~~\mu < 2 \\ {1 \over 8(4-\mu^2)}< \gamma <-{1\over 8\mu^2} ~~,&~~\mu > 2\end{cases}$~~.
\end{itemize}

\begin{figure}[!h]
	\label{fixed2graf}
	\centering
	\includegraphics[height=8cm]{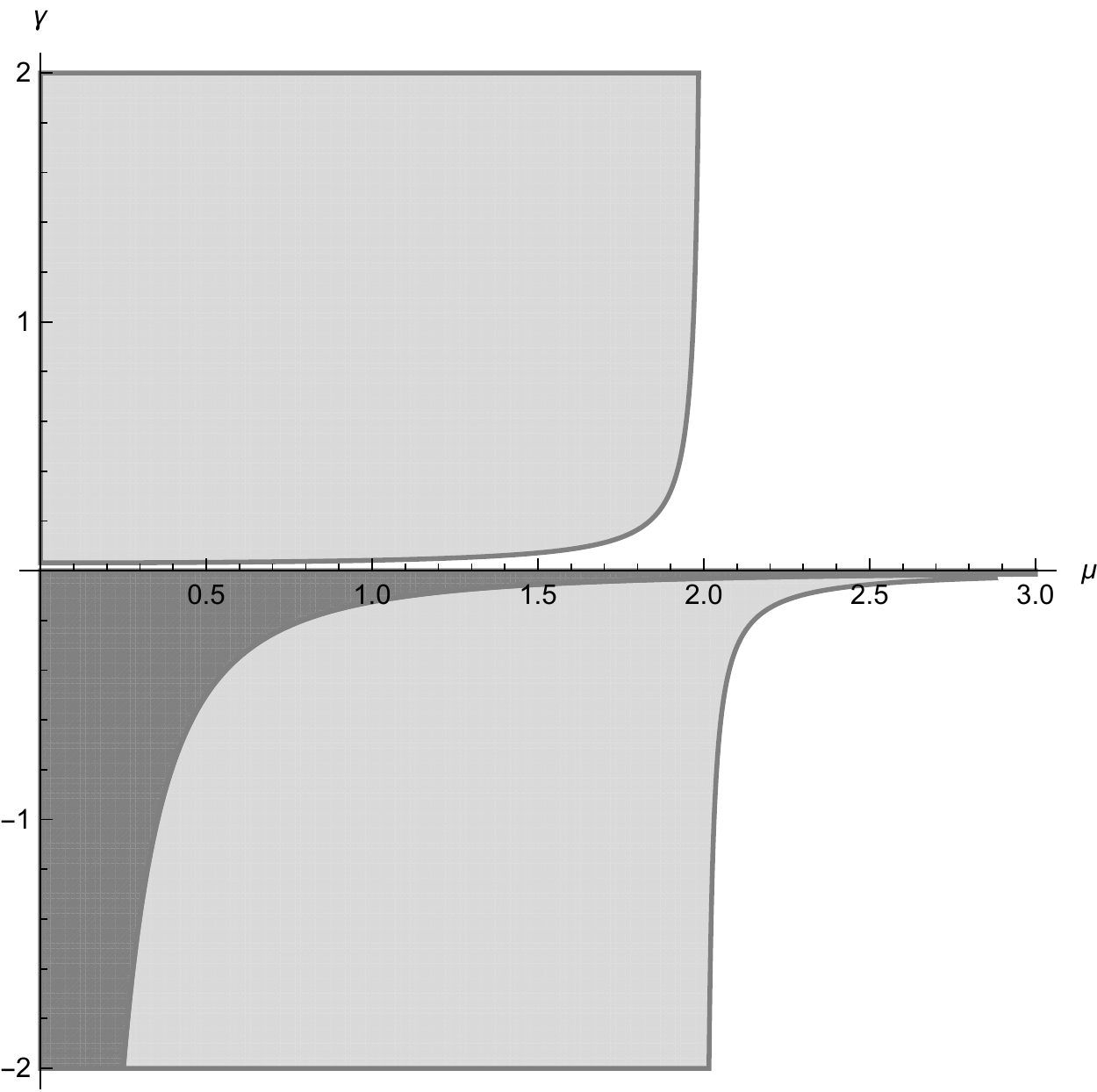}
	\caption{Degenerate eigenvalues of $S_\pm$: real and negative (dark gray), pure imaginary (light gray).}
\end{figure}

Again resorting to Eq. (\ref{zsq}), we find that 

\begin{eqnarray} z\left(\rho,\arccos\left(\pm \sqrt{1 \over 1-{1\over 32\gamma}-{1 \over 4} \mu^2 }\right) \right) = \nonumber \\ {\rho ^2\over (\rho -1)^2 [ 1 + 8 \gamma  \left(\mu ^2-	4\right) -32 \gamma ^2 \rho ^2]}~~,\end{eqnarray}

\noindent so that taking the limit $\rho = 1$ yields

\begin{equation}z^2 = -{1\over 32 \gamma ^2 }~~. \end{equation}

Thus, although the value for $\theta$ is precisely that required to cancel out the divergence obtained in the case of the critical point $N$, we find that it leads to an unphysical, imaginary expansion rate.

\section{\label{sec:levelX} Conclusions}

In this work, we have studied the dynamics of an $SO(3)$-invariant massive vector field \cite{artigo} nonminimally coupled to the curvature.

The resulting system admits De Sitter, exponential inflationary, solutions for a restricted region of the parameter space, $\gamma>1/4$ (cf. Eq. (\ref{action2})). Some specific regimes for exponential inflation were considered; for the  massless case, $\mu=0$, we have obtained physical solutions. The strong coupling limit, $\mu^2\gamma \gg 1$ is only viable if the coupling to the Ricci scalar is stronger than to the Ricci tensor, $\alpha>\beta$. A weak coupling limit, $\mu^2\gamma \ll 1$ is not achievable, as it breaks the aforementioned constraint $\gamma>1/4$. A power law behaviour for the scale factor and vector field was also studied: however, it is not possible to find a simple monomial solution for the dynamical system embodied in Eqs. (\ref{friedmann})-(\ref{motion}).

We studied the dynamical system arising from the equations of motion for this theory, finding 9 finite critical points and 3 critical points at infinity. In the former case, the origin is a trivial critical point, with no impact arising from the nonminimal coupling between the vector field and curvature: the behaviour of this fixed point is thus naturally equivalent to that obtained in Ref. \cite{artigo}. The other 8 non-trivial points lead to a constant expansion rate and are saddle points, with only 2 of them have physical meaning, {\it i.e.} obeying the constraint $\gamma>1/4$.

Regarding the 3 fixed points at infinity, we have $N(1,\pi/2)$ which, depending on the value of $\gamma$, can behave as a saddle point, an unstable point or a focus. If the critical point $N$ is a focus, this leads to a Big Rip scenario, with the Universe evolving towards a infinite expansion rate. The other 2 critical points $S_{\pm}=(1,\pm \frac{8\gamma\sqrt{-\gamma(8\gamma\mu^2+1)}}{1+8\gamma(\mu^2-4)}$) lead to an imaginary expansion rate, and can thus be identified with an oscillating scale factor.

Thus, we conclude that inflationary solutions can be obtained which are driven by a massive vector field, provided the nonminimal coupling to gravity has a non-vanishing effect, a particularly interesting and pleasing new feature of the presented model. We highlight that this requires that the couplings with the Ricci scalar and the Ricci tensor do not cancel each other out, $\alpha \neq \beta $, {\it i.e.} $\gamma \neq 0$.

\end{document}